\newif\ifPDF\PDFtrue
\documentclass{sig-alt-full}
\ifPDF
\usepackage[dvips]{hyperref}
\def\pdfauthor{B. Blanchet, P. Cousot, R. Cousot, Jrme Feret, L. Mauborgne, A. Min, D. Monniaux, X. Rival}%
\def\pdftitle{A Static Analyzer for Large Safety-Critical Software}%
\def\pdfkeywords{Abstract Interpretation; Abstract Domains; Static Analysis; Verification; Floating Point; Embedded, Reactive, Real-Time, Safety-Critical Software.}%
\hypersetup{pdftitle={\pdftitle},pdfauthor={\textcopyright\ \pdfauthor},pdfkeywords={\pdfkeywords}}
\hypersetup{letterpaper}
\hypersetup{raiselinks,pageanchor,plainpages}
\hypersetup{bookmarksopen,bookmarksnumbered}%
\hypersetup{colorlinks,anchorcolor=black,citecolor=black,filecolor=black,linkcolor=black,menucolor=black,pagecolor=black,urlcolor=black}%
\hypersetup{pdfhighlight=/I}
\hypersetup{pdfpagemode={UseOutlines}}
\fi
\renewcommand{\numberofauthors}[1]{\global\aucount=#1
\ifnum\aucount>4\global\originalaucount=\aucount \global\aucount=4\fi 
\global\auskipcount=\aucount\global\advance\auskipcount by 1
\global\multiply\auskipcount by 2
\global\multiply\auskip by \auskipcount
\global\advance\auwidth by -\auskip
\global\divide\auwidth by \aucount
\global\advance\auwidth by 5mm
}
\def\unskipoverpermission{10mm}
\usepackage{comment}
\def\refsection#1{Sect\mbox{.}~\ref{#1}}
\def\refproposition#1{Prop\mbox{.}~\ref{#1}}
\def\refsections#1{Sect\mbox{.}~\refSECTIONS[#1,]}
\def\reffig#1{Fig\mbox{.}~\ref{#1}}
\def\refSECTIONS[#1,#2,#3]{%
\def\vide{}\def\autreslabels{#3}%
\ref{#1}\ifx\autreslabels\vide\ and \ref{#2}\else, \refSECTIONSprime[#2,#3]\fi}
\def\refSECTIONSprime[#1,#2,#3]{%
\def\vide{}\def\autreslabels{#3}%
\ref{#1}\ifx\autreslabels\vide, and \ref{#2}\else, \refSECTIONSprime[#2,#3]\fi}
\usepackage[mathscr]{euscript}
\usepackage{stmaryrd}
\usepackage{url}
\usepackage[latin1]{inputenc}
\usepackage{theorem}
\usepackage{algorithmic}

\renewcommand{\algorithmicelse}{\} \textbf{else} \{}
\renewcommand{\algorithmicendif}{\}}

\newtheorem{proposition}{Proposition}
\theorembodyfont{\rmfamily}

\def\colsep#1{\hskip2\arraycolsep{#1}\hskip2\arraycolsep\ignorespaces}%
\usepackage[nocfg,newitem,alwaysadjust]{paralist}
\newcommand{\abstr}[1]{#1^\sharp}%
\newcommand{\collecting}[1]{#1^{c}}%
\newcommand{\standard}[1]{#1^{s}}%

\newcommand{\semantics}[1]{\llbracket #1 \rrbracket}%
\makeatletter
\def\wideningsymbol{\mathord{\raisebox{0.4ex}{{\fontsize{7pt}{7pt}\selectfont$\bigtriangledown$}}}}
\def\narrowingsymbol{\mathord{\raisebox{0.4ex}{{\fontsize{7pt}{7pt}\selectfont$\bigtriangleup$}}}}
\makeatother
\def\wideningoperator{\ensuremath{\mathbin{\wideningsymbol}}}
\def\narrowingoperator{\ensuremath{\mathbin{\narrowingsymbol}}}
\newcommand{\stagedwidening}[1]{\ensuremath{\mathbin{\wideningsymbol_{\!#1}}}}

\usepackage{color}
\usepackage{graphicx}
\newlength{\todolength}
\gdef\todolist{}
\newif\iftodo\todofalse

\AtEndDocument{\iftodo\typeout{***}\typeout{***}\typeout{*** TODO LIST: \todolist}\typeout{***}\typeout{***}\fi}
\gdef\suggestlist{}
\newif\ifsuggest\suggestfalse
\gdef\nextinsuggestlist{}
\newcommand{\SUGGEST}[3]{%
\suggesttrue
\xdef\suggestlist{\suggestlist\nextinsuggestlist #1 -> #2}\gdef\nextinsuggestlist{, }%
\vspace{1ex}%
\begin{flushleft}
    \color{blue}\fbox{\parbox{\todolength}{\underline{SUGGESTION} [#1 $\rightarrow$ #2]: #3}}
\end{flushleft}%
\vspace{1ex}} 
\AtEndDocument{\ifsuggest\typeout{***}\typeout{***}\typeout{*** 
SUGGESTION LIST: \suggestlist}\typeout{***}\typeout{***}\fi}

\newcommand{\topfloat}{+\infty}

\begin{document}
\conferenceinfo{PLDI'03,}{June 9--11, 2003, San Diego, California, USA.}%
\CopyrightYear{2003}%
\crdata{1-58113-662-5/03/0006}%
\title{A Static Analyzer for Large Safety-Critical
Software
}
\subtitle{(Extended Abstract)%
}

\numberofauthors{4}

\author{
\alignauthor Bruno Blanchet$\,$%
\setcounter{footnote}{0}%
\thanks{\scriptsize~CNRS~(Centre~National~de~la~Recherche~Scientifique)}$\,$
\setcounter{footnote}{3}%
\thanks{\scriptsize~\'Ecole normale sup\'erieure.~{\ttfamily\itshape 
First-name.Last-name}{\ttfamily @ens.fr}}
\alignauthor Patrick Cousot$\,$\smash{$^{\S}$}
\setcounter{footnote}{4}%
\alignauthor Radhia Cousot$\,$\smash{$^{\ast}$}%
\thanks{\scriptsize~\'Ecole polytechnique.~{\ttfamily\itshape 
First-name.Last-name}{\ttfamily 
@polytechnique.fr}\vspace*{-\unskipoverpermission}%
}
\alignauthor J\'{e}r\^{o}me Feret$\,$\smash{$^{\S}$}
  \end{tabular}\\%
  \begin{tabular}[t]{p{\auwidth}}\centering%
   Laurent Mauborgne$\,$\smash{$^{\S}$}
\alignauthor Antoine Min{\'{e}}$\,$\smash{$^{\S}$}
\alignauthor David Monniaux$\,$\smash{$^{\ast}$}\smash{$^{\S}$}
\alignauthor Xavier Rival$\,$\smash{$^\S$}}

\date{\today}
\maketitle

\def\vshrinkspace{2.5}
\begin{abstract}
We show that abstract interpretation-based static program analysis can
be made efficient and precise enough to formally verify a class of
properties for a family of large programs with few or no false alarms. 
This is achieved by refinement of a general purpose static analyzer
and later adaptation to particular programs of the family by the
end-user through parametrization.  This is applied to the proof of
soundness of data manipulation operations at the machine level for
periodic synchronous safety critical embedded software.

The main novelties are the design principle of static analyzers by
refinement and adaptation through parametrization
(\refsections{design-principle,sec:parametrization}),
the symbolic manipulation of
expressions to improve the precision of abstract transfer functions
(\refsection{symbolic}), the octagon (\refsection{octagons}), ellipsoid
(\refsection{sec:ellipsoiddomain}), and decision tree
(\refsection{sec:booleandomain}) abstract domains, all with sound
handling of rounding errors in floating point computations,
widening strategies
(with thresholds: \refsection{staged_widening},
delayed: \refsection{delayed-widening}) and the automatic
determination of the parameters (parametrized packing:
\refsection{packs}).
\end{abstract}
\vspace*{-\vshrinkspace mm}
.
\vspace*{-1mm}
\vspace*{-\vshrinkspace mm}
\vspace*{-\vshrinkspace mm}
\section{Introduction}%
\label{sec:introduction}%
Critical software systems (as found in
industrial plants, automotive, and aerospace applications)
should never fail.  Ensuring that such software does not fail
is usually done by testing, which is expensive
for complex systems with high reliability requirements,
and anyway fails to prove the impossibility of failure.  Formal
methods, such as model checking, theorem proving, and static analysis,
can help.

The definition of ``failure'' itself is difficult, in particular
in the absence of a formal specification.
In this paper, we choose to focus on a particular aspect found in all
specifications for critical software, that is, ensuring that the
critical software never executes an instruction with ``undefined'' or
``fatal error'' behavior, such as out-of-bounds accesses to arrays or
improper arithmetic operations (such as overflows or division
by zero).  Such conditions ensure that the program is written
according to its intended semantics, for example the critical system
will never abort its execution.  These correctness conditions are
automatically extractable from the source code, thus avoiding the need
for a costly formal specification.  Our goal is to prove automatically
that the software never executes such erroneous instructions or, at
least, to give a very small list of program points that may possibly
behave in undesirable ways.

In this paper, we describe our implementation and experimental studies
of static analysis by abstract interpretation over a family of
critical software systems, and we discuss the main technical
choices and possible improvements.

\section{Requirements}%
\label{sec:Requirements}%
When dealing with undecidable questions on program execution, the
verification problem must reconcile \emph{correctness} (which excludes
non exhaustive methods such as simulation or test), \emph{automation}
(which excludes model checking with manual production of a program
model and deductive methods where provers must be manually assisted),
\emph{precision} (which excludes general analyzers which would produce
too many false alarms, i.e., spurious warnings about potential errors),
\emph{scalability} (for software of a few
hundred thousand lines), and \emph{efficiency} (with minimal space and
time requirements allowing for rapid verification during the software
production process which excludes a costly iterative refinement
process).

Industrialized general-purpose static analyzers satisfy all criteria
but precision and efficiency.  Traditionally, static analysis is made
efficient by allowing correct but somewhat imprecise answers to
undecidable questions.  In many usage contexts, imprecision is
acceptable provided all answers are sound and the imprecision rate
remains low (e.g.\ 5 to 15\% of the runtime tests cannot typically be
eliminated).  This is the case for program
optimization (such as static elimination of run-time array bound
checks), program transformation (such as partial evaluation), etc.

In the context of program verification, where human interaction must be
reduced to a strict minimum, false alarms are undesirable.  A 5\% rate
of false alarms on a program of a few hundred thousand lines would
require a several person-year effort to manually prove that no error
is possible.  Fortunately, abstract interpretation theory shows that
for any finite class of programs, it is possible to achieve full
precision and great efficiency \cite{Cousot00-SARA} by discovering an
appropriate abstract domain.  The challenge is to show that this
theoretical result can be made practical by considering infinite but
specific classes of programs and properties to get efficient analyzers
producing few or no false alarms.  A first experiment on smaller
programs of a few thousand lines was quite encouraging
\cite{BlanchetCousotEtAl02-NJ} and the purpose of this paper is
to report on a real-life application showing that the approach does
scale up.

\section{Design Principle}%
\label{design-principle}%
The problem is to find an abstract domain that yields an efficient
and precise static analyzer for the given family of programs.  Our
approach is in two phases, an \emph{initial design} phase by
specialists in charge of designing a parametrizable analyzer followed
by an \emph{adaptation} phase by
end\discretionary{-}{}{-}us\-ers in charge of adapting the analyzer
for (existing and future) programs in the considered family by an
appropriate choice of the parameters of the abstract domain and
the iteration strategy (maybe using some parameter adaptation
strategies provided by the analyser).

\subsection{Initial Design by Refinement}

Starting from an existing analyzer \cite{BlanchetCousotEtAl02-NJ}, the
initial design phase is an iterative manual refinement of the
analyzer.  We have chosen to start from a program in the considered
family that has been running for 10 years without any run-time error,
so that all alarms are, in principle, due to the imprecision of the
analysis.  The analyzer can thus be iteratively refined for this example
until all alarms are eliminated.

Each refinement step starts with a static analysis of the program,
which yields false alarms.
Then a manual backward inspection of the program starting from sample
false alarms leads to the understanding of the origin of the imprecision
of the analysis.  There can be two different reasons for the lack of
precision:
\begin{asparaitem}
\item Some local invariants are expressible in the current version of
the abstract domain but were missed either: 
\begin{asparaitem}
\item because some \emph{abstract transfer function} (\refsection{primitives}) was too coarse, in which case it must be rewritten 
closer to the best abstraction of the concrete transfer function
\cite{CousotCousot79-1}, (\refsection{symbolic});
\item  or because a \emph{widening} (\refsection{widening}) was too 
coarse, in which case the iteration
strategy must be refined (\refsection{parametrizediteration});
\end{asparaitem}
\item Some local invariants are necessary in the correctness proof but
are not expressible in the current version of the abstract domain.  To
express these local invariants, a new abstract domain has to be
designed by specialists and incorporated in the analyzer as an
approximation of the reduced product \cite{CousotCousot79-1} of this
new component with the already existing domain (\refsection{parametrizedabstractdomain}).
\end{asparaitem}

When this new refinement of the analyzer has been implemented, it is
tested on typical examples and then on the full program to verify
that some false alarms have been eliminated.  In general the same
cause of imprecision appears several times in the program; furthermore, one
single cause of imprecision at some program point often leads later to many
false alarms in the code reachable from that program point, so a single
refinement typically eliminates a few dozen if not hundreds of false
alarms.

This process is to be repeated until there is no or very few false
alarms left.

\subsection{Adaptation by Parametrization}

The analyzer can then be used by 
end\discretionary{-}{}{-}us\-ers in charge of proving programs in the
family.  The necessary adaptation of the analyzer to a particular
program in the family is by appropriate choice of some parameters.  An
example provided in the preliminary experience
\cite{BlanchetCousotEtAl02-NJ} was the \emph{widening with
thresholds} (\refsection{staged_widening}). 
Another example is relational domains (such as octagons
\cite{MineAST01}, \refsection{octagons}) which cannot be applied to all global variables
simultaneously because the corresponding analysis would be too
expensive; it is possible to have the user supply
for each program point groups of variables on
which the relational analysis should be independently applied.

In practice we have discovered that the parametrization can be largely
automated (and indeed it is fully automated for octagons as explained
in \refsection{sec:parametrization}).  This way the effort to
manually adapt the analyzer to a particular program in the family is
reduced to a minimum.

\subsection{Analysis of the Alarms}
We implemented and used a \emph{slicer} \cite{Weiser84} 
to help in the alarm inspection process.
If the slicing criterion is an alarm point, the  extracted slice
contains the computations that led to the alarm.
However, the classical data and control dependence-based backward
slicing turned out to yield prohibitively large slices.

In practice we are not interested in the computation of the variables
for which the analyzer already provides a value close to end-user
specifications, and we can consider only the variables we lack
information about (integer or floating point variables that may
contain large values or boolean variables that may take any value
according to the invariant).
In the future we plan to design more adapted forms of
slicing: an {\em abstract slice} would only contain the computations
that lead to an alarm point wherever the
invariant is too weak.

\section{The Considered Family of Programs}

\sloppy

The considered programs in the family are automatically generated
using a proprietary tool from a high-level specification familiar to
control engineers, such as systems of differential equations or
synchronous operator networks (block diagrams as illustrated in
\reffig{fig:blockdiagram}), which is equivalent to
the use of synchronous languages (like
\textsc{Lustre} \cite{HalbwachsCaspiRaymondPilaud91-1}%
).  Such synchronous data-flow specifications are quite
common in real-world safety-critical control systems ranging from
letter sorting machine control to safety control and monitoring
systems for nuclear plants and ``fly-by-wire'' systems.  Periodic
synchronous programming perfectly matches the need for the real-time
integration of differential equations by forward, fixed step numerical
methods.  Periodic synchronous programs have the form:

\begin{algorithmic}%
    \renewcommand{\algorithmicloop}{\textbf{loop forever}}%
    \renewcommand{\algorithmicendloop}{\textbf{end loop}}%
    \STATE \textbf{declare} volatile input, state and output variables;
    \STATE initialize state variables;
    \LOOP
    \STATE \mbox{}\quad{\labelitemii}  read volatile input variables,
    \STATE \mbox{}\quad{\labelitemii}  compute output and state variables,
    \STATE \mbox{}\quad{\labelitemii}  write to volatile output variables;
    \STATE \textbf{wait for next clock tick};
    \ENDLOOP
\end{algorithmic}

\noindent Our analysis proves that no exception can be raised (but the clock
tick) and that all data manipulation operations are sound.  The
bounded execution time of the loop body should also be checked by static
analysis \cite{FerdinandHeckmannEtAL01-EMSOFT} to prove that the
real-time clock interrupt does occur at idle time.

We operate on the C language source code of those systems, ranging
from a few thousand lines to 132,000 lines of C source code (75~kLOC after
preprocessing and simplification as in~\refsection{part:preprocessing}).  We take into account all machine-dependent aspects of
the semantics of C (as described in \cite{BlanchetCousotEtAl02-NJ}) as
well as the periodic synchronous programming aspects (for the
{\textbf{wait}).  We use additional specifications to
describe the material environment with which the software interacts
(essentially ranges of values for a few hardware registers containing
volatile input variables and a maximal execution time to limit the
possible number of iterations in the external
loop\footnote{Most physical systems cannot run forever and some
event counters in their control programs are bounded because of this
physical limitation.}).

The source codes we consider use only a reduced subset of
C, both in the automatically generated glue
code and the handwritten pieces.  As it is often the case with
critical systems, there is no dynamic memory allocation and the use of
pointers is restricted to call-by-reference.  On the other hand, an
important characteristics of those programs is that the number of
global and \texttt{static}\footnote{In C, a
\texttt{static} variable has limited lexical scope yet is persistent
with program lifetime.  Semantically, it is the same
as a global variable with a fresh name.} variables is roughly linear
in the length of the code.  Moreover the analysis must consider the values of
all variables and the abstraction cannot ignore any part
of the program without generating false alarms.  It was therefore a
grand challenge to design an analysis that is precise and does scale
up.

\section{Structure of the Analyzer}

The analyzer is implemented in Objective Caml
\cite{LeroyEtAl01-OCaml}.  It operates in two phases: the
preprocessing and parsing phase followed by the analysis phase.

\subsection{Preprocessing Phase}
\label{part:preprocessing}

The source code is first preprocessed using a standard C preprocessor,
then parsed using a C99-compatible parser.  
Optionally, a simple linker allows programs consisting of several
source files to be processed.

The program is then type-checked and compiled to an intermediate
representation, a simplified version of the abstract syntax tree with
all types explicit and variables given unique identifiers. 
Unsupported constructs are rejected at this point with an error
message.

Syntactically constant expressions are
evaluated and replaced by their value.  Unused global variables are
then deleted.  This phase is important since the analyzed programs
use large arrays representing hardware features with constant
subscripts; those arrays are thus optimized away.

Finally the preprocessing phase includes preparatory work for trace
partitioning (\refsection{tracepartitioning}) and parametrized
packing (\refsection{packs}).

\subsection{Analysis Phase}
\label{sect:iterator}

The analysis phase computes the reachable states in the considered
abstract domain.  This abstraction is formalized by 
a concretization function $\gamma$
\cite{CousotCousot77,CousotCousot79-1,Cousot92-jlc}.  The computation
of the abstraction of the reachable states by the abstract interpreter
is called \emph{abstract execution}.

The abstract interpreter first creates the global and \texttt{static}
variables of the program (the stack-allocated variables are created
and destroyed on-the-fly).  Then the abstract execution is performed
compositionally, by induction on the abstract syntax, and driven by
the \emph{iterator}.

\subsection{General Structure of the Iterator}

The abstract execution starts at a user-supplied entry point for the
program, such as the \texttt{main} function.  Each program construct
is then interpreted by the iterator according to the semantics
of C as well as some information about the target environment (some
orders of evaluation left unspecified by the C norm, the sizes of the
arithmetic types, etc., see \cite{BlanchetCousotEtAl02-NJ}). 
The iterator transforms the C instructions into directives for the
abstract domain that represents the memory
state of the program (\refsection{memoryabstractdomain}), that is, the global,
static and stack-allocated
variables.

The iterator operates in two modes: the \emph{iteration mode} and the
\emph{checking mode}.  The iteration mode is used to
generate invariants; no warning is displayed when some possible errors
are detected.  When in checking mode, the iterator issues a warning
for each operator application that may give an error on the concrete level
(that is to say, the program may be interrupted, such as when dividing by zero,
or the computed result may not obey the end-user specification for this
operator, such as when integers wrap-around due to an overflow).
In all cases, the analysis goes on with the \emph{non-erroneous}
concrete results (overflowing integers are wiped out and not
considered modulo, thus following the end-user intended semantics).

Tracing facilities with various degrees of detail are
also available.  For example the loop invariants which are generated
by the analyzer can be saved for examination.

\subsection{Primitives of the Iterator}
\label{primitives}
Whether in iteration or checking mode, the iterator starts with an
abstract environment $\abstr{E}$ at the beginning of a statement $S$
in the program and outputs an abstract environment
$\abstr{\semantics{S}}(\abstr{E})$ which is a valid abstraction after
execution of statement $S$.  This means that if a concrete environment
maps variables to their values, $\standard{\semantics{S}}$ is a
standard semantics of $S$ (mapping an environment $\rho$ before
executing $S$ to the corresponding environment
$\standard{\semantics{S}}(\rho)$ after execution of $S$),
$\collecting{\semantics{S}}$ is the collecting semantics of $S$
(mapping a set $E$ of environments before executing $S$ to the
corresponding set $\collecting{\semantics{S}}(E) =
\{\standard{\semantics{S}}(\rho)\mid\rho\in E\}$ of environments after
execution of $S$), $\gamma(\abstr{E})$ is the set of concrete
environments before $S$ then $\abstr{\semantics{S}}(\abstr{E})$
over-approximates the set
$\collecting{\semantics{S}}(\gamma(\abstr{E}))$ of environments after
executing $S$ in that $\collecting{\semantics{S}}(\gamma(\abstr{E}))$
$\subseteq$ $\gamma(\abstr{\semantics{S}}(\abstr{E}))$.  The abstract
semantics $\abstr{\semantics{S}}$ is defined as follows:
   
\label{transferfunction}%
\begin{asparaitem}
   \item \emph{Tests}:
  let us consider a conditional\\[1mm]
  \centerline{\begin{tabular}{rcl}
  $S$&=&\mbox{}\textbf{if} (\textit{c}) \texttt{\{} $S_1$ \texttt{\}} 
      \textbf{else} \texttt{\{} $S_2$ \texttt{\}}\\
   \end{tabular}}\\[1mm]
   (an absent
   \textbf{else} branch is considered as an empty execution sequence). 
   The condition $c$ can be assumed to have no side effect and to
   contain no function call, both of which can be handled by first
   performing a program transformation.  The iterator computes:
    \begin{eqnarray*}
        \abstr{\semantics{S}}(\abstr{E})
        &=&
        \abstr{\semantics{S_1}}(\abstr{\textit{guard}}(\abstr{E}, c))
        \mathbin{\abstr{\sqcup}}
        \abstr{\semantics{S_2}}(\abstr{\textit{guard}}(\abstr{E}, \neg c))
    \end{eqnarray*}
 where the abstract domain implements:
    \begin{asparaitem}
    \item $\abstr{\sqcup}$ as the abstract union that is an abstraction
    of the union $\cup$ of sets of environments;
    \item $\abstr{\textit{guard}}(\abstr{E}, c)$ as an approximation of
    $\collecting{\semantics{c}}(\gamma(\abstr{E}))$ where the collecting
    semantics $\collecting{\semantics{c}}(E)=\{\rho\in E\mid
    \standard{\semantics{c}}(\rho)=\mathrm{true}\}$ of the condition $c$
    is the set of concrete environments $\rho$ in $E$ satisfying condition
    $c$.  In practice, the abstract domain only implements
    $\abstr{\textit{guard}}$ for atomic conditions and compound ones are
    handled by structural induction.
\end{asparaitem}
\item \emph{Loops} are by far the most delicate construct to analyze. 
   Let us denote by $\abstr{E}_0$ the environment before the loop:\\[1mm]
  \centerline{\begin{tabular}{l}
            \mbox{}\textbf{while} (\textit{c}) \texttt{\{}
            \textit{body} 
            \texttt{\}}
        \end{tabular}}\\[1mm]
   The abstract loop invariant to be computed for the head of the loop is
   an upper approximation of the least invariant of $F$ where $F(E)$ =
   $\gamma(\abstr{E}_0) \cup \collecting{\semantics{\textit{body}}}
   (\collecting{\semantics{c}}(E))$.
   The fixpoint computation $\abstr{F}(\abstr{E})$ = $\abstr{E}_0$
   $\abstr{\sqcup}$
   $\abstr{\semantics{\textit{body}}}(\abstr{\textit{guard}}(\abstr{E},
   c))$ is always done in iteration mode, requires a widening
   (\refsection{widening}) and stops with an abstract invariant
   $\abstr{E}$ satisfying $\abstr{F}(\abstr{E})$ $\abstr{\sqsubseteq}$
   $\abstr{E}$ (where the abstract partial ordering $x$ $\abstr{\sqsubseteq}$ $y$ implies $\gamma(x)$ 
   $\subseteq$ $\gamma(y)$)
   \cite{Cousot92-jlc}. When in checking mode, the abstract
   loop invariant has first
   to be computed in iteration mode and then, 
   an extra iteration (in checking mode this time),
   starting from this abstract invariant is necessary to collect 
   potential errors.

\item \emph{Sequences} $i_1 \texttt{;} i_2$: first
   $i_1$ is analyzed, then $i_2$, so that:
   \begin{equation*}
    \abstr{\semantics{i_1 \texttt{;} i_2}}(\abstr{E}) =
    \abstr{\semantics{i_2}} \circ 
    \abstr{\semantics{i_1}}(\abstr{E})\enspace.
   \end{equation*}

\item \emph{Function calls} are analyzed by abstract execution of the
function body in the context of the point of call, creating temporary
variables for the parameters and the return value.  Since the considered
programs do not use recursion, this gives a context-sensitive
polyvariant analysis semantically equivalent to inlining.

\item \emph{Assignments} are passed to the abstract domain.

\item \emph{Return statement}:
We implemented the
\texttt{return} statement by carrying over an abstract environment
representing the accumulated return values (and environments, if the
function has side effects).
\end{asparaitem}

\subsection{Least Fixpoint Approximation with Widening and Narrowing}
\label{widening}
The analysis of loops involves the iterative computation of an
invariant $\abstr{E}$ that is such that $\abstr{F}(\abstr{E})
\mathrel{\abstr{\sqsubseteq}} \abstr{E}$ where $\abstr{F}$ is an abstraction of the 
concrete monotonic transfer function $F$ of the test and loop body. 
In abstract domains with infinite height, this is done by
\emph{widening iterations} computing a finite sequence $\abstr{E}_0$ =
$\bot$, \ldots, $\abstr{E}_{n+1}$ = $\abstr{E}_n$ $\wideningoperator$
$\abstr{F}(\abstr{E}_n)$, \ldots, $\abstr{E}_N$ of successive abstract
elements, until finding an invariant $\abstr{E}_N$.  The
\emph{widening operator} \wideningoperator{} should be sound (that is
the concretization of $x$ 
$\wideningoperator$ $y$
should overapproximate the concretizations of $x$ and $y$)
and ensure the termination in
finite time \cite{CousotCousot77,Cousot92-jlc} (see an example in
\refsection{staged_widening}).

In general, this invariant is not the strongest one in the
abstract domain.  This invariant is then made more and more precise by
\emph{narrowing iterations}: $\abstr{E}_N$, \ldots, $\abstr{E}_{n+1}$
= $\abstr{E}_n \narrowingoperator \abstr{F}(\abstr{E}_n)$
where the \emph{narrowing operator} \narrowingoperator{} 
is sound (the concretization of
$x$ $\narrowingoperator$ $y$ is
an upper approximation of the intersection of $x$ and
$y$) and ensures termination \cite{CousotCousot77,Cousot92-jlc}.

\section{Abstract Domains}

The elements of an abstract domain abstract concrete predicates, that
is, properties or sets of program states.  The operations of an
abstract domain are transfer functions abstracting predicate
transformers corresponding to all basic operations in the program
\cite{CousotCousot77}.  The analyzer is fully parametric in the
abstract domain (this is implemented using an Objective Caml functor). 
Presently the analyzer uses the \emph{memory abstract domain}
of \refsection{memoryabstractdomain}, which abstracts sets of program data
states containing data structures such as simple variables, arrays and 
records.  
This abstract domain is itself parametric in the arithmetic
abstract domains (\refsection{sec:atithabsdom}) abstracting
properties of sets of (tuples of) boolean, integer or floating-point
values.  Finally, the precision of the abstract transfer functions
can be significantly improved thanks to symbolic manipulations of the
program expressions preserving the soundness of their abstract semantics
(\refsection{symbolic}).

\subsection{The Memory Abstract Domain}
\label{memoryabstractdomain}

When a C program is executed, all data structures (simple variables, 
arrays, records, etc) are
mapped to a collection of memory cells containing concrete values.
The \emph{memory abstract domain} is an abstraction of sets of such
concrete memory states.  Its elements, called \emph{abstract
environments}, map variables to abstract cells.  The arithmetic
abstract domains operate on the abstract value of one cell for
non-relational ones (\refsection{sec:basic}) and on
several abstract cells for relational ones
(\refsections{octagons,sec:ellipsoiddomain,sec:booleandomain}).  An abstract
value in a abstract cell is therefore the reduction of the abstract
values provided by each different basic abstract domain (that is an
approximation of their reduced product \cite{CousotCousot79-1}).

\subsubsection{Abstract Environments}

An abstract environment is a collection of abstract cells, which
can be of the following four types:
\begin{asparaitem}
    \item An \emph{atomic cell} represents a variable of a simple
    type (enumeration, integer, or float) by an element of the
    arithmetic abstract domain. Enumeration types, including the
    booleans, are considered to be integers.

    \item An \emph{expanded array cell} represents a program array using
    one cell for each element of the array.  
    Formally, let $A=\bigl((v_{1}^{i},\ldots,v_{n}^{i})\bigr)_{i\in\Delta}$
    be the family (indexed by a set $\Delta$) of values of the array 
    (of size $n$) to be abstracted. The abstraction
    is $\bot$ (representing
non\discretionary{-}{}{-}accessibility of dead code) when $A$ is empty.
Otherwise the abstraction
    is an abstract array $\abstr{A}_{e}$ of size $n$ such the expanded array cell
    $\abstr{A}_{e}[k]$ is the abstraction of $\bigcup_{i\in\Delta}v_{k}^{i}$
    for $k=1$, \ldots, $n$. Therefore the abstraction is component-wise,
    each element of the array being abstracted separately.
    \item A \emph{shrunk array cell} represents a program array using a
    single cell.  Formally the abstraction is a shrunk array cell
    $\abstr{A}_{s}$ abstracting
    $\bigcup_{k=1}^{n}\bigcup_{i\in\Delta}v_{k}^{i}$.
    All elements of the array are thus ``shrunk'' together.  We use this
    representation for large arrays where all that matters is the range of
    the stored data.

    \item A \emph{record cell} represents a program record
    (\texttt{struct}) using one cell for each field of the record.
    Thus our abstraction is field-sensitive.
\end{asparaitem}

\subsubsection{Fast Implementation of Abstract Environments}
\label{part:zapped_maps}

A naive implementation of abstract environments may use an array.  We
experimented with in-place and functional arrays and found this
approach very slow.  The main reason is that abstract union
$\abstr{\sqcup}$ 
operations are
expensive, because they operate in time linear in the number of
abstract cells; since both the number of global variables (whence of
abstract cells) and the number of tests (involving the abstract
union $\abstr{\sqcup}$) are linear in the length of the code, this
yields a quadratic time behavior.

A simple yet interesting remark is that in most cases, abstract
union
operations are applied between abstract environments that
are identical on almost all abstract cells: branches of tests 
modify a few abstract cells only. It is therefore desirable that those
operations should have a complexity proportional to the number of
\emph{differing} cells between both abstract environments. We chose 
to implement
abstract environments using \emph{functional maps} implemented as sharable
balanced
binary trees, with short-cut evaluation when computing the abstract
union, abstract intersection, widening or narrowing of physically identical subtrees
\cite[§6.2]{BlanchetCousotEtAl02-NJ}. 
An additional benefit of sharing is that it contributes to the rather
light memory consumption of our analyzer.  

On a 10,000-line example we
tried \cite{BlanchetCousotEtAl02-NJ}, the execution time was divided
by seven, and we are confident that the execution times would have
been prohibitive for the longer examples.
The efficiency of functional maps in the context of
sophisticated static analyses has also been observed by
\cite{ManevichEtAl-SAS02} for representing first-order structures.

\subsubsection{Operations on Abstract Environments}

Operations on a C data structure are translated into operations on
cells of the current abstract environments.  Most translations are
straightforward.
\begin{asparadesc}
\item[{\labelitemii}\ Assignments:]
In general, an assignment $\mathit{lvalue} := e$ is translated into the
assignment of the abstract value of $e$ into the abstract cell
corresponding to $\mathit{lvalue}$. However, for array assignments,
such as $x[i] := e$, one has to note that the array index $i$ may not be
fully known, so all cells possibly corresponding to $x[i]$ may either be
assigned the value of $e$, or keep their old value. In the
analysis, these cells are assigned the upper bound of their old
abstract value and the abstract value of $e$. Similarly, for a shrunk
array $x$, after an assignment $x[i] := e$, the cell representing $x$
may contain either its old value (for array elements not modified by the
assignment), or the value of $e$.
\item[{\labelitemii}\ Guard:]\label{guard}%
The translation of concrete to abstract guards is not detailed
since similar to the above case of assignments. 

\item[{\labelitemii}\ Abstract union, widening, narrowing:] Performed
cell\discretionary{-}{}{-}wise between abstract environments.
\end{asparadesc}

\subsection{Arithmetic Abstract Domains}
\label{sec:atithabsdom}
\begingroup

The non-relational arithmetic abstract domains abstract
sets of numbers while the relational domains abstract sets of tuples
of numbers.  The basic abstract domains we started with
\cite{BlanchetCousotEtAl02-NJ} are the intervals and the clocked
abstract domain abstracting time.  They had to be significantly
refined using octagons (\refsection{octagons}), ellipsoids
(\refsection{sec:ellipsoiddomain}) and decision trees
(\refsection{sec:booleandomain}).

\subsubsection{Basic Abstract Domains}
\label{sec:basic}
\begin{asparaitem}

\item  \emph{The Interval Abstract Domain.}
The first, and simplest, implemented domain is the domain of
intervals, for both integer and floating-point values
\cite{CousotCousot77}.  Special care has to be taken in the case of
floating-point values and operations to always perform rounding in the right
direction and to handle special IEEE \cite{FloatIEEE} values such as
infinities and \emph{NaN}s (Not a Number).

\medskip

\item \emph{The Clocked Abstract Domain.}
A simple analysis using the intervals gives a large number of false
warnings.  A great number of those warnings originate from possible
overflows in counters triggered by external events.  Such errors
cannot happen in practice, because those events are counted at most once
per clock cycle, and the number of clock cycles in a single execution
is bounded by the maximal continuous operating time of the system.

We therefore designed a parametric abstract domain. (In our case, 
the parameter is the interval domain~\cite{BlanchetCousotEtAl02-NJ}.)
Let $\abstr{X}$ be an abstract domain for a single scalar variable.
The elements of the clocked domain consist in triples in
$({\abstr{X}})^3$. A triple $(\abstr{v},\abstr{v}_-,\abstr{v}_+)$
represents the set of values $x$ such that $x \in \gamma(\abstr{v})$,
$x - \textit{clock} \in \gamma(\abstr{v}_-)$ and
$x + \textit{clock} \in \gamma(\abstr{v}_+)$, where \textit{clock} is a
special, hidden variable incremented each time the analyzed program
waits for the next clock signal.
\end{asparaitem}

\subsubsection{The Octagon Abstract Domain}
\label{octagons}

Consider the following program fragment:%

\begin{algorithmic}
\STATE\mbox{}\qquad \texttt{R} := \texttt{X}$-$\texttt{Z};
\STATE\mbox{}\qquad \texttt{L} := \texttt{X};
\STATE\mbox{}\qquad \textbf{if}\ {(\texttt{R}$>$\texttt{V})}\ \ \texttt{L} := \texttt{Z}$+$\texttt{V};
\end{algorithmic}

\noindent At the end of this fragment, we have $\texttt{L}\leq\texttt{X}$.
In order to prove this, the analyzer must discover that, when the test
is true, we have $\texttt{R}=\texttt{X}-\texttt{Z}$ and
$\texttt{R}>\texttt{V}$, and deduce from this that
$\texttt{Z}+\texttt{V}<\texttt{X}$ (up to rounding).
This is possible only with a \emph{relational domain} able to capture 
simple linear inequalities between variables.

Several such domains have been proposed, such as the widespread polyhedron
domain \cite{CousotHalbwachs78}.
In our prototype, we have chosen the recently developed 
\emph{octagon abstract domain} \cite{octagonlib,MineAST01}, 
which is less precise 
but faster than the polyhedron domain: it can represent sets of
constraints of the form $\pm x \pm y \leq c$, and its complexity is
cubic in time and quadratic in space (w.r.t. the number of variables),
instead of exponential for polyhedra.
Even with this reduced cost, the huge number of live variables prevents
us from representing sets of concrete environments as one big abstract state
(as it was done for polyhedra in \cite{CousotHalbwachs78}). Therefore
we partition the set of variables into small subsets and use one octagon for 
some of these subsets (such a group of variables being then called
a \emph{pack}).
The set of packs is a parameter of the analysis which can be determined 
automatically (\refsection{octagon-packs}).

Another reason for choosing octagons is the lack of support for 
floating-point arithmetics in the polyhedron domain.
Designing relational domains for floating-point variables is indeed
a difficult task, not much studied until recently \cite{Martel02}.
On one hand, the abstract domain must be sound with respect to 
the concrete floating-point semantics (handling rounding, \emph{NaN}s, etc.); 
on the other hand it should use floating-point numbers internally to manipulate
abstract data for the sake of efficiency.
Because invariant manipulations in relational domains
rely on some properties of the real field not
true for floating-points (such as $x+y\leq c$ and $z-y\leq d$ implies
$x+z\leq c+d$), it is natural to consider that abstract values represent
subsets of $\mathbb{R}^N$ (in the relational invariant $x+y\leq c$,
the addition $+$ is considered in $\mathbb{R}$,
without rounding, overflow, etc.).
Our solution separates the problem in two.
First, we design a sound abstract domain for variables in the real field
(our prototype uses the \emph{octagon library} \cite{octagonlib}
which implementation is described in \cite{MinePADO01}).
This is much easier for octagons than for polyhedra, as most computations
are simple (addition, multiplication and division by $2$).
Then, each floating-point expression is transformed into a sound
approximate real expression taking rounding, overflow, etc. into
account (we use the linear forms described in
\refsection{symbolic}) and evaluated by the abstract domain.

Coming back to our example, it may seem that octagons are not expressive enough
to find the correct invariant as
$\texttt{Z}+\texttt{V}<\texttt{X}$ is not representable in an octagon.
However, our assignment transfer function is smart enough to extract
from the environment the interval $[c,d]$ where \texttt{V} ranges 
(with $d\leq \texttt{R}_{M}$ where $\texttt{R}_{M}$ is an upper
bound of  $\texttt{R}$ already computed by the analysis) and
synthesize the invariant $c\leq\texttt{L}-\texttt{Z}\leq d$, which is
sufficient to prove that subsequent operations on \texttt{L} will not
overflow.
Thus, there was no need for this family of programs to use a more expressive
and costly relational domain.

Remark that this approach provides a generic way of implementing
relational abstract domains on floating-point numbers. It is parametrized
by:
\begin{asparaitem}
\item  a strategy for the determination of packs (\refsection{octagon-packs});
\item  an underlying abstract domain working in the real field.
\end{asparaitem}
Aspects specific to floating-point computation (such as rounding and
illegal operations) are automatically taken care of by our approach.

\subsubsection{The Ellipsoid Abstract Domain}
\label{sec:ellipsoiddomain}

To achieve the necessary precision, several new abstract domains had
to be designed.  We illustrate the general approach on the case
of the ellipsoid abstract domain.  

By inspection of the parts of
the program on which the previously described analyses provide no
information at all on the values of some program variables, we
identified code of the form:
\begin{algorithmic}
\IF {({\texttt{B}})}
\STATE \texttt{Y} := $i$;
\STATE \texttt{X} := $j$; 
\ELSE 
\STATE $\texttt{X}'$ := $a$\texttt{X} $-$ $b$\texttt{Y} $+$ $t$;
\STATE \texttt{Y} := \texttt{X};
\STATE \texttt{X} := $\texttt{X}'$;
\ENDIF  
\end{algorithmic}

\noindent where $a$ and $b$ are floating-point constants, $i$, $j$ and
$t$ are floating-point expressions, $\texttt{B}$ is a boolean
expression, and $\texttt{X}$, $\texttt{X}'$, and $\texttt{Y}$ are
program variables.  The previously described analyses yield the
imprecise result that $\texttt{X}$ and $\texttt{Y}$ may take any
value.  This apparently specialized style of code is indeed quite
frequent in control systems since it implements the simplified second
order digital filtering discrete-time system illustrated in
\reffig{fig:blockdiagram}.
\begin{figure}
\includegraphics[scale=0.875]{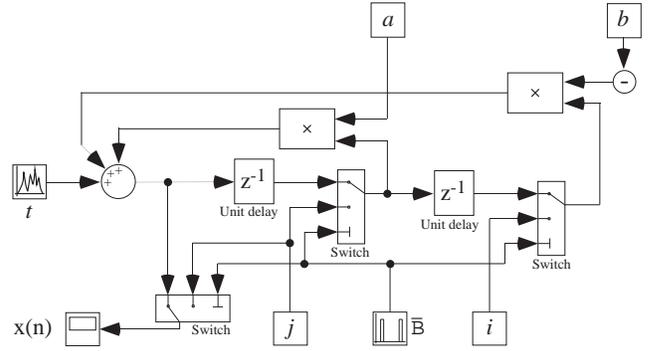}
\caption{A simplified second-order digital filtering system.}
\label{fig:blockdiagram}
\end{figure}

The first branch is a reinitialization step, the second branch
consists in an affine transformation $\Phi$.  Since this code is
repeated inside loops, the analysis has to find an invariant preserved
by this code.  We looked manually for such an invariant on typical
examples, identified the above generic form (essentially depending on
$a$ and $b$), then designed a generic abstract domain
$\varepsilon_{a,b}$ able to discover such invariants, implemented the
abstract domain lattice and transfer operations and finally let
the analyzer automatically instantiate the specific analysis to the
code (in particular to parts that may not have been inspected).

To find an interval that contains the values of $\texttt{X}$ and
$\texttt{Y}$ in the specific case where we can compute bounds to the
expression $t$ by the previously described analyses, say $|t| \leq
t_M$, we have designed a new abstract domain $\varepsilon_{a,b}$ based
on ellipsoids, that can capture the required invariant.  More
precisely, we can show that:
\begin{proposition}\label{prop:ellipseinvariant}
If $0 < b < 1$, $a^2 - 4b < 0$, and $k \geq \left(\frac{t_M}{1-\sqrt{b}}\right)^2$, 
then the constraint $\mathtt{X}^2 - a\mathtt{X}\mathtt{Y} + b\mathtt{Y}^2 \leq k$ is
preserved by the affine transformation $\Phi$.
\end{proposition}
The proof of this proposition follows by algebraic manipulations using
standard linear algebra techniques.  In our examples, the conditions
on $a$ and $b$ required in \refproposition{prop:ellipseinvariant} are
satisfied.  We still have to design the abstract operations to
propagate the invariant in the program, and to take into account
rounding errors that occur in floating-point computations (and are not
modeled in the above proposition).

Having fixed two floating-point numbers $a$ and $b$ such that
$0 < b < 1$ and $a^2 - 4b < 0$, we 
present a domain $\varepsilon_{a,b}$, for describing  sets of
ellipsoidal constraints.
An element in $\varepsilon_{a,b}$ is a function $r$ which 
maps a pair of variables $(\texttt{X},\texttt{Y})$ to a floating-point
number $r(\mathtt{X}, \mathtt{Y})$ such that
$\texttt{X}^2-a\texttt{X}\texttt{Y}+b\texttt{Y}^2\leq
r(\texttt{X},\texttt{Y})$.

We briefly describe some primitives and transfer functions of our domain:
\begin{asparaitem}
\item \emph{Assignments.} Let $r\in\varepsilon_{a,b}$ be the abstract
  element describing some constraints before a statement
  $\texttt{X} := e$, our goal is to compute the abstract element $r'$
  describing a set of constraints satisfied after this statement:
\begin{enumerate}
\item in case $e$ is a variable $\texttt{Y}$, 
each constraint containing $\texttt{Y}$ gives a constraint for
$\texttt{X}$. Formally,
we take $r'$ such that $r'(U,V)=r(\sigma U,\sigma V)$ where $\sigma$ is
the substitution of the variable $\texttt{Y}$ for the
variable $\texttt{X}$;
\item in case $e$ is an expression of the form
  $a\texttt{Y}+b\texttt{Z}+t$, we first remove any
  constraint containing $\texttt{X}$, then we add a new constraint
  for $\texttt{X}$ and $\texttt{Y}$. We therefore take:
\[r'=r[(\mathtt{X},\_) \mapsto \topfloat][(\_,\mathtt{X}) \mapsto \topfloat][(\mathtt{X},\mathtt{Y})\mapsto \delta(r(\mathtt{Y},\mathtt{Z}))]\,.\]
We have used the function $\delta$ defined as follows:
\newcommand{\era}{\left(4f\frac{|a|\sqrt{b}+b}{\sqrt{4b-a^2}}\right)}
 \newcommand{\erb}{(1+f)t_M}
{\[\delta(k) =\left(\left(\sqrt{b}+\era\right)\sqrt{k}+\erb\right)^2\]}%
 where $f$ is the greatest relative error of a float with respect to a
 real and $t \in [ -t_M, t_M]$.  Indeed, we can show that, if $\mathtt{Y}^2
 - a\mathtt{Y}\mathtt{Z} + b\mathtt{Z}^2 \leq k$ and $\mathtt{X} = a\mathtt{Y} - b\mathtt{Z} + t$, then in exact real arithmetic $\mathtt{X}^2 - a\mathtt{X}\mathtt{Y} + b\mathtt{Y}^2
 \leq (\sqrt{bk} + t_M)^2$, and taking into account rounding errors, we
 get the above formula for $\delta(k)$;
      
\item otherwise, we remove all constraints containing 
$\texttt{X}$ by taking 
$r'$ = $r[(\texttt{X},\_)\mapsto \topfloat][(\_,\texttt{X}) \mapsto \topfloat]$\footnote{~This is also the case for initialization.}.
\end{enumerate}
\item \emph{Guards} are ignored, i.e.,\ $r'$ = $r$.
\item \emph{Abstract union}, \emph{intersection}, \emph{widening} and
\emph{narrowing} are computed component-wise.  The widening uses
thresholds as described in \refsection{staged_widening}.
\end{asparaitem}

The abstract domain ${\varepsilon}_{a,b}$ cannot compute accurate results by 
itself, mainly because of inaccurate assignments (in case 3.) and
guards. Hence we use an approximate 
reduced product with the interval domain.
A reduction step consists in substituting in the function $r$ 
the image of a couple $(\texttt{X},\texttt{Y})$ by the smallest 
element among $r(\texttt{X},\texttt{Y})$ and the floating-point number
$k$ such that $k$ is the 
least upper bound to the evaluation 
of the expression $\texttt{X}^2-a\texttt{X}\texttt{Y}+b\texttt{Y}^2$ 
in the floating-point numbers when considering the computed interval 
constraints.
In case the values of the variable $\texttt{X}$ and $\texttt{Y}$ are proved to be equal, we can be much more precise and take the smallest element among $r(\texttt{X},\texttt{Y})$ and the 
least upper bound to the evaluation of 
the expression $(1-a+b)\texttt{X}^2$.

These reduction steps are performed:
\begin{asparaitem}

\item before computing the union between two abstract elements $r_1$
and $r_2$, we reduce each constraint $r_i(\texttt{X},\texttt{Y})$ such
that $r_i(\texttt{X},\texttt{Y})=\topfloat$ and
$r_{3-i}(\texttt{X},\texttt{Y})\not=\topfloat$ (where $i\in\{1;2\}$);

\item before computing the widening between two abstract elements
$r_1$ and $r_2$, we reduce each constraint
$r_2(\texttt{X},\texttt{Y})$ such that
$r_2(\texttt{X},\texttt{Y})=\topfloat$ and
$r_{1}(\texttt{X},\texttt{Y})\not=\topfloat$;

\item before an assignment of the form
$\texttt{X}' := a\texttt{X}-b\texttt{Y}+t$, we refine
the constraints $r(\texttt{X},\texttt{Y})$.

\end{asparaitem}
These reduction steps are especially useful in handling a reinitialization iteration.

Ellipsoidal  constraints are then used to reduce the intervals of
variables: after each assignment $A$ of the form
$\texttt{X}' := a\texttt{X}-b\texttt{Y}+t$, we use the fact that 
$|\texttt{X}'| \leq
2\sqrt{b}\sqrt{\frac{r'(\texttt{X}',\texttt{X})}{4b-a^2}}$, where $r'$
is the abstract element describing a set of ellipsoidal constraints just after 
the assignment $A$.

This approach is generic and has been applied to handle the digital
filters in the program.

\subsubsection{The Decision Tree Abstract Domain}%
\label{sec:booleandomain}%
Apart from numerical variables, the code uses also a great deal of
boolean values, and no classical numerical domain deals precisely
enough with booleans. In particular, booleans can be used in the
control flow and we need to relate the value of the booleans to some
numerical variables. Here is an example:

\begin{algorithmic}
\STATE\mbox{}\qquad \texttt{B} :=  (\texttt{X}=0);
\STATE\mbox{}\qquad \textbf{if} {($\neg$ \texttt{B})}\ \ \texttt{Y} := 1/\texttt{X}; 
\end{algorithmic}

\noindent
We found also more complex examples where a numerical variable could 
depend on whether a boolean value had changed or not.
In order to deal precisely with those examples, we implemented a simple
relational domain consisting in a decision tree with leaf an 
arithmetic abstract domain\footnote{The arithmetic abstract domain is generic. In 
practice, the interval domain was sufficient.}. The decision trees 
are reduced by ordering boolean variables (as in \cite{bryant86}) and
by performing some opportunistic sharing of subtrees.

The only problem with this approach is that the size of decision trees
can be exponential in the number of boolean variables, and the code contains
thousands of global ones. So we extracted a set of variable packs, and
related the variables in the packs only, as explained 
in \refsection{boolean-packs}.

\subsection{Symbolic Manipulation of Expressions}
\label{symbolic}

We observed, in particular for non-relational abstract domains, that
transfer functions proceeding by structural induction on expressions
are not precise when the variables in the expression are not
independent.
Consider, for instance, the simple assignment
$\texttt{X} := \texttt{X}-0.2*\texttt{X}$ performed in the interval domain
in the environment $\texttt{X}\in [0,1]$.
Bottom-up evaluation will give
$\texttt{X}-0.2*\texttt{X}\Rightarrow [0,1]-0.2*[0,1]\Rightarrow[0,1]-[0,0.2]\Rightarrow [-0.2,1]$.
However, because the same $\texttt{X}$ is used on both sides of the $-$ operator, the precise result should have been $[0,0.8]$.

In order to solve this problem, we perform some simple algebraic 
simplifications on expressions before feeding them to the abstract domain.
Our approach is to \emph{linearize} each expression $\mathbf{e}$, that is to say,
transform it into a linear form $\ell\llbracket\mathbf{e}\rrbracket$ on the set of variables
$v_1,\ldots,v_{N}$ with interval coefficients:
$\ell\llbracket\mathbf{e}\rrbracket
=\sum_{i=1}^N [\alpha_i,\beta_i]v_i+[\alpha,\beta]$. The linear form
$\ell\llbracket\mathbf{e}\rrbracket$ is computed by recurrence on the structure of
$\mathbf{e}$.
Linear operators on linear forms 
(addition, subtraction, multiplication and division by
a constant interval) are straightforward.
For instance, $\ell\llbracket \texttt{X}-0.2*\texttt{X}\rrbracket=0.8*\texttt{X}$, which will be evaluated
to $[0,0.8]$ in the interval domain.
Non-linear operators (multiplication of two linear forms, division by a linear
form, non-arithmetic operators) 
are dealt by evaluating one or both linear form argument into an interval.

Although the above symbolic manipulation is correct in the real field,
it does not match the semantics of C expressions for two reasons:
\begin{asparaitem}
\item floating-point computations incur rounding;
\item errors (division by zero, overflow, etc.) may occur.
\end{asparaitem}

Thankfully, the systems we consider conform to the IEEE 754 norm
\cite{FloatIEEE} that describes rounding very well (so that, e.g., the
compiler should be prevent from using the \emph{multiply-add-fused
instruction} on machines
for which the result of a multiply-add computation may be slightly
different from the floating point operation operation $A + (B \times
C)$ for some input values $A$, $B$, $C$).
Thus, it is easy to modify the recursive construction of linear forms from
expressions to add the error contribution for each operator.
It can be an \emph{absolute} error interval, or a \emph{relative} error 
expressed as a linear form.
We have chosen the absolute error
which is more easily implemented and turned out to be precise enough.

To address the second problem, we first evaluate the expression in the
abstract interval domain and proceed with the linearization 
to refine the result only if no possible arithmetic error was reported.
We are then guaranteed that the simplified linear form has the same semantics
as the initial expression.

\endgroup

\section{Adaptation via Parametrization}
\label{sec:parametrization}

In order to adapt the analyzer to a particular program of the
considered family, it may be necessary to provide information to help
the analysis.  A classical idea is to have users provide assertions
(which can be proved to be invariants and therefore ultimately
suppressed).  Another idea is to use parametrized abstract domains in
the static program analyzer.  Then the static analysis can be adapted
to a particular program by an appropriate choice of the parameters. 
We provide several examples in this section.  Moreover we show how
the analyzer itself can be used in order to help or even automatize
the appropriate choice of these parameters.

\subsection{Parametrized Iteration Strategies}
\label{parametrizediteration}

\subsubsection{Loop Unrolling}
\label{loopunrolling}
In many cases, the analysis of loops is made more precise by treating
the first iteration of the loop separately from the following ones;
this is simply a semantic \emph{loop unrolling} transformation: a
\emph{while} loop may be expanded as follows:\\[1mm]
    \centerline{\begin{tabular}{l}
            \mbox{}\textbf{if} (\textit{condition}) \texttt{\{} 
            \textit{body}; 
            \textbf{while} (\textit{condition}) 
            \texttt{\{} 
            \textit{body} 
            \texttt{\}} 
            \mbox{}\texttt{\}}
        \end{tabular}}\\[1mm]
The above transformation can be iterated $n$ times, where the concerned
loops and the unrolling factor $n$ are user-defined parameters.  In
general, the larger the $n$, the more precise the analysis, and the
longer the analysis time.

\subsubsection{Widening with Thresholds}\label{staged_widening}

Compared to normal interval analysis \cite[§2.1.2]{Cousot92}, ours
does not jump straight away to $\pm\infty$, but goes through a
number of thresholds.
The \emph{widening with thresholds} \stagedwidening{T} for the interval
analysis of \refsection{sec:basic} is parametrized by a \emph{threshold set}
$T$ that is a finite set of numbers containing $-\infty$ and $+\infty$
and defined such that:
\begin{displaymath}
    [a,b]\stagedwidening{T}[a',b']=\begin{array}[t]{@{}l@{}}
    [\mathit{if}\:a'<a \:\mathit{then}\:\max\{\ell\in T\mid \ell \leq
    a'\}\:\mathit{else}\:a,\\
    \mbox{}\quad\mathit{if}\: b'>b \:\mathit{then}\:\min\{h\in T\mid h
    \geq b'\}\:\mathit{else}\:b]
    \end{array}
\end{displaymath}

In order to illustrate the benefits of this parametrization (see others
in \cite{BlanchetCousotEtAl02-NJ}), let $x_{0}$ be the initial
value of a variable \texttt{X} subject to assignments of the form
$\mathtt{X} := \alpha_{i}\mathbin{\mathtt{\ast}}\mathtt{X}+\beta_{i}$,
$i\in\Delta$ in the main loop, where the $\alpha_{i}$, $\beta_{i}$,
$i\in\Delta$ are floating point constants such that
$0\leq\alpha_{i}<1$.  Let be any $M$ such that
$M\geq\max\{|x_{0}|,\frac{|\beta_{i}|}{1- \alpha_{i}}, i\in\Delta\}$. 
We have $M\geq|x_{0}|$ and $M\geq\alpha_{i}M+|\beta_{i}|$ and
so all possible sequences $x^{0}=x_{0}$, $x^{n+1}=\alpha_{i}
x^{n}+\beta_{i}$, $i\in\Delta$ of values of variable \texttt{X} are
bounded since $\forall n\geq0:|x^{n}|\leq M$.  Discovering $M$ may be
difficult in particular if the constants $\alpha_{i}$, $\beta_{i}$,
$i\in\Delta$ depend on complex boolean conditions.  
As long as the set $T$ of thresholds contains some number greater or
equal to the minimum $M$, the interval analysis of \texttt{X} with
thresholds $T$ will prove that the value of \texttt{X} is bounded at
run-time since some element of $T$ will be an admissible $M$. 

In practice we have chosen $T$ to be $(\pm\alpha\lambda^k)_{0 \leq k \leq
N}$. The choice of $\alpha$ and $\lambda$ mostly did not matter much
in the first experiments. After the analysis had been well refined and
many causes of imprecision removed, we had to choose a smaller value
for $\lambda$ to remove some false alarms. In any case,
$\alpha\lambda^N$ should be large enough; otherwise, many false alarms
for overflow are produced.

\newpage
\subsubsection{Delayed Widening}
\label{delayed-widening}

When widening the previous iterate by the result of the transfer
function on that iterate at each step as in \refsection{widening},
some values which can become stable after two steps of widening may
not stabilize.  Consider the example:

\begin{algorithmic}
\STATE \texttt{X} := \texttt{Y} + $\gamma$;
\STATE \texttt{Y} := $\alpha$ $\ast$ \texttt{X} + $\delta$
\end{algorithmic}

\noindent This should be equivalent to \texttt{Y} := $\alpha$ $\ast$ \texttt{Y} + $\beta$
(with $\beta=\delta + \alpha \gamma$), and so a widening with thresholds should
find a stable interval. But if we perform a widening with thresholds at
each step, each time we widen \texttt{Y}, \texttt{X} is increased to a value
surpassing the threshold for \texttt{Y}, and so \texttt{X} is widened
to the next stage, which in turn increases \texttt{Y} further and the
next widening stage increases the value of \texttt{Y}. This
eventually results in top abstract values for \texttt{X} and \texttt{Y}.

In practice, we first do $N_0$ iterations with unions on all abstract domains, 
then we do widenings unless a variable which was not stable becomes
stable (this is the case of \texttt{Y} here when the threshold is big enough
as described in \refsection{staged_widening}). We add a fairness
condition to avoid livelocks in cases for each iteration there exists
a variable that becomes stable.

\subsubsection{Floating Iteration Perturbation}
\label{floating-iteration-perturbation}

The stabilization check for loops considered in
\refsection{primitives} has to be adjusted because of the floating
point computations in the abstract.  Let us consider that $[a,b]$ is
the mathematical interval of values of a variable $\mathtt{X}$ on
entry of a loop.  We let $F_{\mathbb{C},\mathbb{A}}([a,b])$ be the
mathematical interval of values of $\mathtt{X}$ after a loop iteration. 
$\mathbb{C}=\mathbb{R}$ means that the concrete operations in the loop
are considered to be on mathematical real numbers while
$\mathbb{C}=\mathbb{F}$ means that the concrete operations in the loop
are considered to be on machine floating point numbers.  If
$F_{\mathbb{R},\mathbb{A}}([a,b])$ = $[a',b']$ then
$F_{\mathbb{F},\mathbb{A}}([a,b])$ =
$[a'-\epsilon_{1},b'+\epsilon_{1}]$ because of the cumulated concrete
rounding errors $\epsilon_{1}\geq 0$ when evaluating the loop
body\footnote{~We take the rounding error on the lower and
upper bound to be the same for simplicity.}.  The same way $\mathbb{A}=\mathbb{R}$ means that
the interval abstract domain is defined ideally using mathematical
real numbers while $\mathbb{A}=\mathbb{F}$ means that the interval
abstract domain is implemented with floating point operations
performing rounding in the right direction.  Again, if
$F_{\mathbb{C},\mathbb{R}}([a,b])$ = $[a',b']$ then
$F_{\mathbb{C},\mathbb{F}}([a,b])$ =
$[a'-\epsilon_{2},b'+\epsilon_{2}]$ because of the cumulated abstract
rounding errors during the static analysis of the loop body.  The analyzer
might use $F_{\mathbb{F},\mathbb{F}}$ which is sound since if
$F_{\mathbb{R},\mathbb{R}}([a,b])$ = $[a',b']$ then
$F_{\mathbb{F},\mathbb{F}}([a,b])$ =
$[a'-\epsilon_{1}-\epsilon_{2},b'+\epsilon_{1}+\epsilon_{2}]$ which takes
both the concrete and abstract rounding errors
into account (respectively $\epsilon_{1}$ and $\epsilon_{2}$).

\smallskip

Mathematically, a loop invariant for variable $\mathtt{X}$ is an interval $[a,b]$ 
such that  $F_{\mathbb{F},\mathbb{R}}([a,b])\subseteq[a,b]$.  However,
the loop stabilization check is made as
$F_{\mathbb{F},\mathbb{F}}([a,b])\subseteq[a,b]$, which is sound but
incomplete: if $F_{\mathbb{F},\mathbb{R}}([a,b])$ is very close to
$[a,b]$, e.g.\ $F_{\mathbb{F},\mathbb{R}}([a,b])$ = $[a,b]$ then,
unfortunately, $F_{\mathbb{F},\mathbb{F}}([a,b])$ =
$[a-\epsilon_{2},b+\epsilon_{2}]\nsubseteq[a,b]$. This will
launch useless additional iterations whence a loss of time and precision.

\smallskip

The solution we have chosen is to overapproximate
$F_{\mathbb{F},\mathbb{F}}$ by $\widehat{F}_{\mathbb{F},\mathbb{F}}$
such that $\widehat{F}_{\mathbb{F},\mathbb{F}}([a,b])$ = $[a'-\epsilon*|a'|,b'+\epsilon*|b'|]$ where $[a',b']$ = $F_{\mathbb{F},\mathbb{F}}([a,b])$ and
$\epsilon$ is a parameter of the analyzer chosen to be an upper bound
of the possible abstract rounding errors in the program loops.
Then the loop invariant interval is computed iteratively with
$\widehat{F}_{\mathbb{F},\mathbb{F}}$, which is simply less precise
than with $F_{\mathbb{F},\mathbb{F}}$, but sound.  
The loop stabilization test is performed with
$F_{\mathbb{F},\mathbb{F}}$ which is sound.  It is also more precise
in case $\epsilon*(\textit{min}\{|a'|;|b'|\})$ is greater than the
absolute error on the computation of
$F_{\mathbb{F},\mathbb{F}}([a'-\epsilon*|a'|,b'+\epsilon*|b'|])$.  We have not
investigated about the existence (nor about the automatic computation)
of such a parameter in the general case yet, nevertheless
attractiveness of the encountered fixpoints made the chosen parameter
convenient.

\subsubsection{Trace Partitioning}
\label{tracepartitioning}
In the abstract execution of the program, when a test is met,
both branches are executed and then the abstract environments computed
by each branch are merged.
As described in \cite{BlanchetCousotEtAl02-NJ} we can get a more
precise analysis by delaying this merging.

This means that:\\
  \centerline{\begin{tabular}{l}
            \mbox{}\textbf{if} (\textit{c}) \texttt{\{} 
            $S_1$ 
            \texttt{\}} 
            \textbf{else} \texttt{\{} 
            $S_2$ 
            \texttt{\}} 
            \textit{rest}
        \end{tabular}}\\[1mm]
is analyzed as if it were\\[1mm]
  \centerline{\begin{tabular}{l}
            \mbox{}\textbf{if} (\textit{c}) \texttt{\{} 
             $S_1$;\ 
            \textit{rest}
             \texttt{\}} 
             \textbf{else} \texttt{\{} 
            $S_2$;\ 
            \textit{rest}
            \mbox{}\texttt{\}}\enspace.
         \end{tabular}}\\[1mm]
A similar technique holds for the unrolled iterations of loops.

As this process is quite costly, the analyzer performs this \emph{trace
partitioning} in a few end-user selected functions, and the traces are
merged at the return point of the function. Informally, in our case, the
functions that need partitioning are those iterating simultaneously
over arrays \mbox{\texttt{a[]}}
and \mbox{\texttt{b[]}}  such that \mbox{\texttt{a[$i$]}}
and \mbox{\texttt{b[$i$]}} are linked by an important numerical constraint
which does not hold in general for \mbox{\texttt{a[$i$]}} and
\mbox{\texttt{b[$j$]}} where $i \neq j$. This solution was simpler
than adding complex invariants to the abstract domain.

\subsection{Parametrized Abstract Domains}%
\label{parametrizedabstractdomain}%

\label{packs}%

Recall that our relational domains (octagons of \refsection{octagons}, and
decision trees of \refsection{sec:booleandomain}) 
operate on small packs of variables for efficiency reasons.
This packing is determined syntactically before the analysis.
The packing strategy is a parameter of the analysis; it gives a
trade-off between accuracy (more, bigger packs) and speed
(fewer, smaller packs).
The strategy must also be adapted to the family of programs to be analyzed.

\subsubsection{Packing for Octagons}%
\label{octagon-packs}%

We determine a set of packs of variables and use one octagon for each
pack.
Packs are determined once and for all, before the analysis starts, 
by examining variables that interact in linear assignments within small 
syntactic blocks (curly-brace delimited blocks).
One variable may appear in several packs and we could do some
information propagation (i.e.\ \emph{reduction}~\cite{CousotCousot79-1})
between octagons at analysis time, using common variables as pivots;
however, this precision gain was not needed in our experiments.
There is a great number of packs, but each pack is small; it is our guess
that our packing strategy constructs, for our program family,
a linear number of constant-sized octagons, effectively resulting in
a cost linear in the size of the program.
Moreover, the octagon packs are efficiently manipulated using functional 
maps, as explained in \refsection{part:zapped_maps},
to achieve sub-linear time costs \emph{via} sharing of unmodified octagons.

Our current strategy is to create one pack for each syntactic block in the 
source code and put in the pack all variables that appear in a linear 
assignment or test within the associated block, ignoring what happens in
sub-blocks of the block.
For example, on a program of 75~kLOC,
2,600 octagons were detected, each containing four variables on average.
Larger packs (resulting in increased cost and precision) could be created by
considering variables appearing in one or more levels of nested blocks; 
however, we found that, in our program family, it does not improve
precision.

\subsubsection{Packing Optimization for Octagons}%
\label{octagon-packing-optimization}%

Our analyzer outputs, as part of the result, whether each octagon actually
improved the precision of the analysis.
It is then possible to re-run the analysis using only packs that were proven
useful, thus greatly reducing the cost of the analysis.
(In our 75~kLOC example, only 400 out of the 2,600 original
octagons were in fact useful.)
Even when the program or the analysis parameters are modified, it is
perfectly safe to use a list of useful packs output by a
previous analysis.
We experimented successfully with the following method:
generate at night an up-to-date list of good octagons by a full, lengthy 
analysis and work the following day using this list to cut analysis costs.

\subsubsection{Packing for Decision Trees}%
\label{boolean-packs}%

In order to determine useful packs for the decision trees of
\refsection{sec:booleandomain}, we used the following strategy: each time a 
numerical variable 
assignment depends on a boolean, or a boolean assignment depends on
a numerical variable, we put both variables in a tentative pack.
If, later, we find a program point where the numerical variable is inside
a branch depending on the boolean, we mark the pack as confirmed. 
In order to deal with complex boolean dependences, if we find
an assignment \texttt{b} := \textit{expr} where \textit{expr}
is a boolean expression,
we add \texttt{b} to all packs containing a variable in \textit{expr}.
In the end, we just keep the confirmed packs.

At first, we restrained the boolean expressions used to extend the
packs to simple boolean variables (we just considered
\texttt{b} := \texttt{b'}) and the packs contained at most four boolean
variables and dozens of false alarms were removed. But we discovered
that more false alarms could be removed if we extended those assignments
to more general expressions. The problem was that packs could then
contain up to 36 boolean variables, which gave very bad performance. So
we added a parameter to restrict arbitrarily the number of boolean
variables in a pack. Setting this parameter to three yields an efficient and
precise analysis of boolean behavior.

\section{Experimental Results}
\begin{figure}
\includegraphics{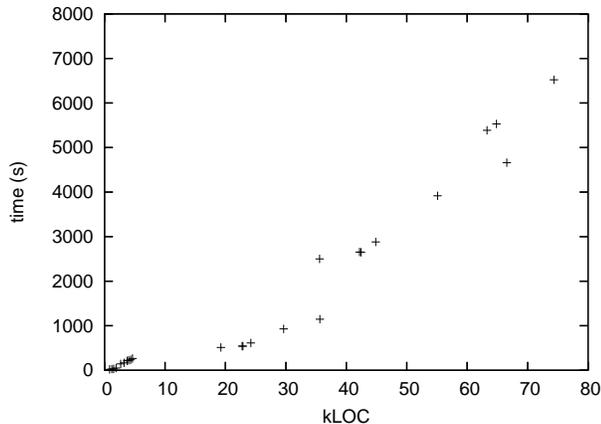}\vspace{-5mm}
\caption{Total analysis time for the family of programs \underline{without}
packing optimization (\refsection{octagon-packing-optimization}).}
\label{fig:total-time}
\end{figure}

The main program we are interested in is 132,000 lines of C with
macros (75~kLOC after preprocessing and simplification as
in~\refsection{part:preprocessing}) and has about 10,000 global/static
variables (over 21,000 after array expansion as in
\refsection{memoryabstractdomain}).  We had 1,200 false alarms with
the analyzer \cite{BlanchetCousotEtAl02-NJ} we started with. 
The refinements of the analyzer described in this paper reduce the
number of alarms down to 11 (and even 3, depending on the versions of
the analyzed program).
\reffig{fig:total-time} gives the total analysis time for a family of
related programs on commodity hardware (2.4~GHz, 1~Gb RAM PC), using
a slow but precise iteration strategy.

The memory consumption of the analyzer is reasonable (550~Mb for the
full-sized program). Several parameters, for instance the size of the
octagon packs (\refsection{octagon-packs}), allow for a space-precision
trade-off.

The packing optimization strategy of reusing
results from preceding analysis to reduce the number of octagons
(\refsection{octagon-packing-optimization}) reduces, on the largest example
code, memory consumption from 550~Mb to 150~Mb and time from
1\,h~40\,min to 40\,min. Furthermore, the current automatic tuning of
the iteration strategy may be made more efficient, using fewer
iterations and thus reducing analysis time.

\section{Related Work}

Let us discuss some other verification methods that could have been
considered.  Dynamic checking methods were excluded for a safety
critical system (at best data can be collected at runtime and checked
offline).  Static methods requiring compiler or code instrumentation
(such as\ \cite{EnglerEtAl-OSDI2000}) were also excluded in our
experiment since the certified compiler as well as the compiled code,
once certified by traditional methods, cannot be modified without
costly re-certification processes.  Therefore we only consider the
automated static proof of software run-time properties, which has been
a recurrent research subject since a few decades.

\subsection{Software Model Checking}%
\label{software-model-checking}%

Software model checking \cite{Holzmann07-SE} has proved very
useful to trace logical design errors, which in our case has already
been performed at earlier stages of the software development, whereas
we concentrate on abstruse machine implementation aspects of the
software.  Building a faithful model of the program (e.g.\ in
\textsc{Promela} for \textsc{Spin} \cite{Holzmann07-SE}) would be just
too hard (it can take significantly more time to write a model than it
did to write the code) and error-prone (by checking a manual
abstraction of the code rather than the code itself, it is easy to
miss errors).  Moreover the abstract model would have to be designed
with a finite state space small enough to be fully explored (in the
context of verification, not just debugging), which is very difficult
in our case since sharp data properties must be taken into account. 
So it seems important to have the abstract model automatically
generated by the verification process, which is the case of the
abstract semantics in static analyzers.

\subsection{Dataflow Analysis and Software Abstract Model Checking}%
\label{dataflow-analyzers}%

Dataflow analyzers (such as ESP \cite{DasLernerSeigle-PLDI02})
as well as abstraction based software model checkers (such as
a.o.\ \textsc{Blast} \cite{HenzingerEtAlPOPL02}, CMC
\cite{EnglerEtAl-OSDI2002} and \textsc{Slam}
\cite{BallRajamani-POPL02}) have made large inroads in tackling
programs of comparable size and complexity.  Their impressive
performance is obtained thanks to coarse abstractions (e.g.\ 
resulting from a program ``shrinking'' preprocessing phase
\cite{DasLernerSeigle-PLDI02,AdamsBallEtAl-SAS02} or obtained by a
globally coarse but locally precise abstraction
\cite{EnglerEtAl-OSDI2002}). In certain cases, the abstract model is
just a finite automaton, whose transitions are triggered by certain
constructions in the source code \cite{EnglerEtAl-OSDI2000}; this
allow checking at the source code level high-level properties, such as
``allocated blocks of memory are freed only once'' or ``interrupts are
always unmasked after being blocked'', ignoring dependencies on data.

The benefit of this coarse abstraction
is that only a small part of the program control and/or data have to
be considered in the actual verification process.  This idea did not
work out in our experiment since merging paths or data inevitably
leads to many false alarms.  On the contrary we had to resort to
context-sensitive polyvariant analyses (\refsection{primitives}) with
loop unrolling (\refsection{loopunrolling}) so that the size of the
(semantically) ``expanded'' code to analyze is much larger than that
of the original code. Furthermore, the properties we prove include fine
numerical constraints, which excludes simple abstract models.

\subsection{Deductive Methods}%
\label{deductive-methods}%

Proof assistants (such as \textsc{Coq} \cite{COQ}, ESC
\cite{FlanaganEtAl-PLDI02} or PVS
\cite{OwreShankarStringer-Calvert:FM-Trends98}) face semantic problems
when dealing with real-life programming
languages.  First, the prover has to take the
machine-level semantics into account (e.g., floating-point arithmetic
with rounding errors as opposed to real numbers, which is far from
being routinely available \footnote{For example ESC is simply unsound
with respect to modular arithmetics \cite{FlanaganEtAl-PLDI02}.}). 
Obviously, any technique for analyzing machine arithmetic will face
the same semantic problems.  However, if the task of taking concrete
and rounding errors into account turned out to be feasible for
our automated analyzer, this task is likely to be daunting in
the case of complex decision procedures operating on ideal arithmetic
\cite{OwreShankarStringer-Calvert:FM-Trends98}. Furthermore, exposing
to the user the complexity brought by those errors is likely to make
assisted manual proof harrowing.

A second semantic difficulty is that the prover needs to operate on
the C source code, not on some model written in a prototyping language
so that the concrete program semantics must be incorporated in the prover (at
least in the verification condition generator).
Theoretically, it is possible to do a ``deep embedding'' of the
analyzed program into the logic of the proof assistant --- that is,
providing a mathematical object describing the syntactic structure of the
program as well as a formal semantics of the programming language.
Proving any interesting property is then likely to be extremely
difficult.
``Shallow embeddings'' --- mapping the original program to a
corresponding ``program'' in the input syntax of the prover --- are
easier to deal with, but may be difficult to produce in the presence
of nondeterministic inputs, floating-point rounding errors etc\dots

The last and main
difficulty with proof assistants is that they must be assisted, in
particular to help providing inductive arguments (e.g.\ invariants). 
Of course these provers could integrate abstract domains in the form
of abstraction procedures (to perform online abstractions of arbitrary
predicates into their abstract form) as well as decision procedures
(e.g.\ to check for abstract inclusion $\sqsubseteq^\sharp$).  The
main problem is to have the user provide program independent hints,
specifying when and where these abstraction and decision procedures
must be applied, as well as how the inductive arguments can be
discovered, e.g.\ by iterative fixpoint approximation, without
ultimately amounting to the implementation of a static program
analysis.

Additionally, our analyzer is designed to run on a whole family of
software, requiring minimal adaptation to each individual program. In
most proof assistants, it is difficult to change the program without
having to do a considerable amount of work to adapt proofs.

\subsection{Predicate Abstraction}%
\label{predicate-abstraction}%

\emph{Predicate abstraction}, which consists in specifying an
abstraction by providing the atomic elements of the abstract domain in
logical form \cite{GrafSaidi-CAV97} e.g.\ by representing sets of
states as boolean formulas over a set of base predicates, would
certainly have been the best candidate.  Moreover most implementations
incorporate an automatic refinement process by success and failure
\cite{BallEtAl-PLDI01,HenzingerEtAlPOPL02} whereas we successively
refined our abstract domains manually, by experimentation.  In
addition to the semantic problems shared by proof assistants, a number
of difficulties seem to be insurmountable to automate this design
process in the present state of the art of deductive methods:

\subsubsection{State Explosion Problem:}%
\label{state-explosion-problem}%

To get an idea of the size of the
necessary state space, we have dumped the main loop invariant (a
textual file over 4.5~Mb).  

The main loop invariant includes 6,900 boolean interval assertions
($x\in[0,1]$), 9,600 interval assertions 
($x\in[a,b]$), 25,400 clock assertions (\refsection{sec:basic}), 
19,100 additive octagonal assertions ($a \leq x + y \leq b$), 19,200
subtractive octagonal assertions ($a \leq x - y \leq b$, see
\refsection{octagons}), 100 decision trees (\refsection{sec:booleandomain}) 
and 1,900
ellipsoidal assertions (\refsection{sec:ellipsoiddomain})\footnote{Figures are rounded to the
closest hundred.  We get more assertions than variables because in the
10,000 global variables arrays are counted once whereas the element-wise
abstraction yields assertions on each array element.  Boolean
assertions are needed since booleans are integers in C.}.

In order to allow for the reuse of boolean model checkers, the
conjunction of true atomic predicates is usually encoded as a boolean
vector over boolean variables associated to each predicate
\cite{GrafSaidi-CAV97} (the disjunctive completion
\cite{CousotCousot79-1} of this abstract domain can also be used to
get more precision \cite{BallEtAl-PLDI01,HenzingerEtAlPOPL02}, but
this would introduce an extra exponential factor).  Model checking
state graphs corresponding to several tenths of thousands of boolean
variables (not counting hundreds of thousands of program points) is still a
real challenge.  Moreover very simple static program analyzes, such as
Kildall's constant propagation \cite{Kildall73-1}, involve an infinite abstract domain
which cannot be encoded using finite boolean vectors thus requiring the
user to provide beforehand all predicates that will be indispensable to
the static analysis (for example the above mentioned loop invariant
involves, e.g., over 16,000 floating point constants at most 550 of
them appearing in the program text).

Obviously some of the atomic predicates automatically generated by our
analysis might be superfluous.  On one hand it is hard to say which
ones and on the other hand this does not count all other predicates
that may be indispensable at some program point to be locally precise. 
Another approach would consist in trying to verify each potential
faulty operation separately (e.g., focus on one instruction that may
overflow at a time) and generate the abstractions lazily
\cite{HenzingerEtAlPOPL02}.  Even though repeating this analysis over
100,000 times might be tractable, the real difficulty is to
automatically refine the abstract predicates (e.g.\ to discover that
considered in \refproposition{prop:ellipseinvariant}).

\subsubsection{Predicate Refinement:}%
\label{predicate-refinement}%

Predicate abstraction \emph{per
se} uses a finite domain and is therefore provably less powerful than
our use of infinite abstract domains (see \cite{CousotCousot-PLILP92},
the intuition is that all inductive assertions have to be provided
manually).  Therefore predicate abstraction is often accompanied by a
refinement process to cope with false alarms
\cite{BallEtAl-PLDI01,HenzingerEtAlPOPL02}.

Under specific conditions, this refinement can be proved equivalent to
the use of an infinite abstract domain with widening
\cite{BallEtAl-TACAS02}.  

Formally this refinement is a fixpoint computation
\cite{Cousot00-SARA, GiacobazziQuintarelli-SAS01} at the concrete
semantics level, whence introduces new elements in the abstract domain
state by state without termination guarantee whereas, e.g., when
introducing clocks from intervals or ellipsoids from octagons we
exactly look for an opposite more synthetic point of view.  Therefore
the main difficulty of counterexample-based refinement is still to
automate the presently purely intellectual process of designing
precise and efficient abstract domains.

\section{Conclusion}
\vskip3pt
In this experiment, we had to cope with stringent
requirements. Industrial constraints prevented us from requiring any
change in the production chain of the code. For instance, it was
impossible to suggest changes to the library functions that would
offer the same functionality but would make the code easier to
analyze. Furthermore, the code was mostly automatically generated from
a high-level specification that we could not have access to, following
rules of separation of design and verification meant to prevent the
intrusion of unproved high-level assumptions into the verification
assumptions. It was therefore impossible to analyze the high-level
specification instead of analyzing the C code.

That the code was automatically generated had contrary effects.  On
the one hand, the code fit into some narrow subclass of the whole C
language.  On the other hand, it used some idioms not commonly found
in human-generated code that may make the analysis more difficult; for
instance, where a human would have written a single test with a
boolean connective, the generated code would make one test, store the
result into a boolean variable, do something else do the second test
and then retrieve the result of the first test.  Also, the code
maintains a considerable number of state variables, a large number of
these with local scope but unlimited lifetime.  The interactions
between several components are rather complex since the considered
program implement complex feedback loops across many interacting
components.

Despite those difficulties, we developed an analyzer with a very high
precision rate, yet operating with reasonable computational power and
time.  Our main effort was to discover an appropriate abstraction
 which we did by manual refinement
through experimentation of an existing analyzer
\cite{BlanchetCousotEtAl02-NJ} and can be later adapted by end-users
to particular programs through parametrization
(\refsections{symbolic,sec:parametrization}).  To achieve this, we
had to develop two specialized abstract domains
(\refsections{sec:ellipsoiddomain,sec:booleandomain}) and improve an
existing domain (\refsection{octagons}).  

The central idea in this approach is that once the analyzer has been
developed by specialists, end-users can adapt it to other programs in
the family without much efforts.  However coming up with a tool that
is effective in the hands of end users with minimal expertise in
program analysis is hard.  This is why we have left to the user the
simpler parametrizations only (such as widening thresholds in
\refsection{staged_widening} easily found in the program
documentation) and automated the more complex ones (such as
parametrized packing \refsection{packs}).  Therefore, the approach
should be economically viable.

\balancecolumns
\end{document}